\documentstyle[aps,psfig,graphpap,graphics]{revtex}
\begin{document}
\title{Periodic waves in bimodal optical fibers}
\author{K. W. Chow$^1$ \thanks{
e-mail: kwchow@hkusua.hku.hk}, K. Nakkeeran$^2$ \thanks{
e-mail: ennaks@polyu.edu.hk}, and Boris A. Malomed$^3$ \thanks{
e-mail: malomed@post.tau.ac.il}}
\address{$^1$Department of Mechanical Engineering, University of Hong Kong, \\
Pokfulam Road, Hong Kong}
\address{$^2$Photonics Research Center and Department of Electronic and \\
Information Engineering, The Hong Kong Polytechnic University, \\
Hung Hom, Kowloon, Hong Kong}
\address{$^3$Department of Interdisciplinary Studies, Faculty of Engineering, \\
Tel Aviv University, Tel Aviv 69978, Israel}
\maketitle

\begin{abstract}
We consider coupled nonlinear Schr\"odinger equations (CNLSE) which govern
the propagation of nonlinear waves in bimodal optical fibers. The nonlinear
transform of a dual-frequency signal is used to generate an
ultra-short-pulse train. To predict the energy and width of pulses in the
train, we derive three new types of travelling periodic-wave solutions,
using the Hirota bilinear method. We also show that all the previously
reported periodic wave solutions of CNLSE can be derived in a systematic
way, using the Hirota method.

\end{abstract}


%\newpage

\section{Introduction}

Steady increase of bit rates provided by fiber-optic telecommunication
networks is necessary to support the rapid growth of the data traffic. The
development of high-repetition-rate generators of high-quality ultra-short
pulses plays a crucial role for the increase of transmission capacity in
fiber-optic networks, as well as for optical-signal processing \cite
{yamamoto00}. Since available opto-electronic direct-modulation techniques
cannot provide for the generation of pulses at a repetition rate higher than
40 GHz, nonlinear transformation of a dual-frequency signal in an optical
fiber is an attractive method for applications to ultrahigh-speed
telecommunications. Indeed, the repetition rate of the generated pulses can
be tuned by adjusting frequencies of the two input waves. A nonlinear
technique based on the modulational instability of a pump wave induced by a
small periodic signal was first suggested by Hasegawa \cite{hasegawa84} and
realized experimentally by Tai et al. \cite{tai86}. Another nonlinear method
is based on reshaping of a beat signal which results from superposition of
two pump waves of equal power. Nonlinear reshaping of the beat signal into
well-separated short pulses was previously demonstrated by several methods,
using dispersion decreasing fibers (DDF) \cite{mamyshev91}, intra-pulse
stimulated Raman scattering in DDF \cite{chernikov93}, or switching in a
nonlinear fiber-loop mirror \cite{chernikov93_2}. Other techniques based on
step-like and comb-like dispersion-profiled fibers, composed of segments of
conventional uniform fibers with different dispersions, have also been
proposed for the generation of picosecond and sub-picosecond pulse trains
with high repetition rates \cite
{chernikov93_3,chernikov94,chernikov94_2,chernikov96,swanson95}. However,
the common feature of these techniques, employed for the reshaping of a beat
signal, is that they rely on careful control of dispersion distribution in
complex fiber systems. An alternative approach, based on multiple four-wave
mixing (FWM) in an optical fiber with {\it constant} anomalous dispersion,
has been demonstrated theoretically for the generation of compressed pulse
trains from a dual-frequency pump field \cite{trillo94}. In particular, the
optimum fiber length (propagation distance at which the pulse width attains
a minimum) for given values of input beat-signal parameters, has been
accurately predicted by calculating the propagation distance at which the
maximum energy conversion from the pumps into the nearest FWM-sidebands
occurs \cite{trillo94}. Experimental demonstration of the generation of a
transform-limited, pedestal-free pulse train with high repetition rate by
means of this technique of the multi-wave-mixing temporal compression in the
standard non-zero dispersion-shifted fiber was reported recently \cite
{pitois02}.

For this kind of ultra-short pulse-train generation from the beating of
two-mode signals, one must investigate periodic-wave solutions of nonlinear
equations governing the fiber system. The single-mode wave propagation in
fibers is governed by the well-known nonlinear Schr\"{o}dinger (NLS)
equation \cite{HAS-KOD,GPA},

\[
\psi _{z}-i\beta \psi _{tt}+i|\psi |^{2}\psi =0, 
\]
where $\psi $ is the slowly varying envelope of the axial electrical field
and $2\beta $ is the group-velocity-dispersion (GVD) coefficient, while $z$
and $t$ are spatial and temporal variables. The normal and anomalous GVD
corresponds, respectively, to $\beta >0$ and $\beta <0$.

Single-mode fibers actually support co-propagation of two different modes
with orthogonal polarizations, which, in the case of anomalous GVD and
without birefringence, is described by coupled NLS equations (CNLSE) \cite
{MAN,Hiet,men,pka}.

\begin{eqnarray}
i\psi _{1z}-\beta \psi _{1tt}+\kappa \psi _{2}+\left( |\psi _{1}|^{2}+\sigma
|\psi _{2}|^{2}\right) \psi _{1} &=&0,  \nonumber \\
i\psi _{2z}-\beta \psi _{2tt}+\kappa \psi _{1}+\left( |\psi _{2}|^{2}+\sigma
|\psi _{1}|^{2}\right) \psi _{2} &=&0,  \label{cnlse}
\end{eqnarray}

\noindent where $\sigma >0$ is the relative cross-phase modulation (XPM)
coefficient, and the linear coupling coefficient $\kappa $ accounts for
possible twist of the fiber in the case when $\psi _{1}$ and $\psi _{2}$
represent orthogonal linear polarizations, or elliptic deformation of the
fiber's cross section if $\psi _{1}$ and $\psi _{2}$ correspond to circular
polarizations. In fact, the linear coupling will play a crucially important
role in the present work.

Complete integrability of the system (\ref{cnlse}) by means of the inverse
scattering transform (IST) in the case $\sigma =1,\,\kappa =0$ was shown by
Manakov \cite{MAN}. In a completely integrable system, one can use IST to
construct not only solitons, but periodic solutions too. However, $\sigma
\neq 1$ in real optical fibers, hence Eqs. (\ref{cnlse}) are not integrable.
In particular, $\sigma =2$ for two modes with circular polarizations, as
well as for a different situation, when two signals carried by different
wavelengths co-propagate in the fiber. For two linearly polarized modes, one
has $\sigma =2/3$, and in the more general case of elliptically polarized
modes, the XPM coefficient may take any value from $2/3<\sigma <2$. All
these physical situations come under the umbrella of the non-integrable
CNLSE system.

The objective of the present paper is to find new periodic-wave solutions of
the CNLSE system (\ref{cnlse}) for both anomalous-GVD and normal-GVD
regimes. Solutions will be obtained by dint of the Hirota method (HM) \cite
{hir} through the Jacobi's elliptic functions. Links to previously found
periodic-wave solutions of the same system will be established too.

The rest of the paper is organized as follows. In section 2, we produce
three new types of periodic-wave solutions for the anomalous-GVD case, and
two new solution types for the normal-GVD\ case. In section 3, we briefly
consider the soliton (long-wave) limit of these solutions, concluding that
they amount to anti-dark and dark (generally, gray) solitons in the
anomalous- and normal-GVD cases, respectively. Conclusions are formulated in
section 4, and some technicalities are collected in two Appendices.

\section{Periodic waves in the CNLSE system}

\subsection{The anomalous-dispersion regime}

In this subsection we consider the case $\beta <0$ in Eqs. (\ref{cnlse}),
adopting the normalization $\beta =-1$. Some periodic-wave solutions of
nonintegrable CNLSE systems have been found earlier for this case in Refs. 
\cite{flo_trem_pla,kos_uzu_oc,flo_trem_oc}. Here we shall obtain three new
kinds of periodic solutions by means of HM, in terms of the theta and Jacobi
elliptic functions. We shall also show how previously known periodic
solutions of the system (\ref{cnlse}) can be readily found by means of HM.

We start with the transformation

\begin{equation}
\psi _{1}=\frac{1}{2}\left( a+ib\right) ,\,\,\psi _{2}=\frac{1}{2}\left(
a-ib\right) ,  \label{ab}
\end{equation}

\noindent which casts Eqs. (\ref{cnlse}) with $\beta =-1$ in the form

\begin{eqnarray}
ia_{z}+a_{tt}+\kappa a+\frac{1}{2}(|a|^{2}+|b|^{2})a+\frac{1}{4}(\sigma
-1)(a^{2}+b^{2})a^{\star } &=&0,  \nonumber \\
ib_{z}+b_{tt}-\kappa b+\frac{1}{2}(|a|^{2}+|b|^{2})b+\frac{1}{4}(\sigma
-1)(a^{2}+b^{2})b^{\star } &=&0,  \label{cnlsem}
\end{eqnarray}
where the asterisk stands for the complex conjugation. Stationary solutions
to these equations are sought for in the form

\begin{eqnarray*}
a=\xi (t)\exp (-i\Omega z), \hspace{0.25in} b=\eta (t)\exp (-i\Omega z),
\end{eqnarray*}

\noindent where the functions $\xi $ and $\eta $ are assumed real. Coupled
equations for these functions are tantamount to the stationary version of
the Manakov system,

\begin{eqnarray}
\xi _{xx}+(\Omega +\kappa )\xi +\frac{1}{4}(1+\sigma )(\xi ^{2}+\eta
^{2})\xi &=&0,  \nonumber \\
\eta _{xx}+(\Omega -\kappa )\eta +\frac{1}{4}(1+\sigma )(\xi ^{2}+\eta
^{2})\eta &=&0.  \label{odes}
\end{eqnarray}
Despite the asymmetry of equations (\ref{odes}), accounted for by the terms
proportional to $\pm \kappa $, this system is an integrable one (it has a
second dynamical invariant, besides the Hamiltonian) \cite{Hiet}.

We now employ the Hirota bilinear operator:

\begin{equation}
D_{x}^{m}D_{t}^{n}g\cdot f=\left. \left( \frac{\partial }{\partial x}-\frac{
\partial }{\partial x^{\prime }}\right) ^{m}\left( \frac{\partial }{\partial
t}-\frac{\partial }{\partial t^{\prime }}\right) ^{n}g(x,t)f(x^{\prime
},t^{\prime })\right| _{x=x^{\prime },t=t^{\prime }},  \label{hir-der}
\end{equation}
to rewrite Eqs. (\ref{odes}) in the trilinear form \cite{kwc1}:

\[
\xi \equiv \frac{g}{f},\,\eta \equiv \frac{G}{f}, 
\]

\begin{eqnarray}
f(D_{x}^{2}+\Omega +\kappa )g\cdot f+g\left[ -D_{x}^{2}f\cdot f+\frac{1}{4}
(1+\sigma )(g^{2}+G^{2})\right] &=&0,  \label{trans3} \\
f(D_{x}^{2}+\Omega -\kappa )G\cdot f+G\left[ -D_{x}^{2}f\cdot f+\frac{1}{4}
(1+\sigma )(g^{2}+G^{2})\right] &=&0.  \nonumber
\end{eqnarray}
Periodic waves are obtained by choosing $g$, $G$, and $f$ as products of
theta functions. The Hirota bilinear forms of the theta functions can be
simplified by using known properties of these functions. Technical details
are omitted in the main text, as the method is very similar to that
presented in Refs. \cite{kwc1,kwc2}; a brief sketch of the derivation of the
solutions displayed below is given in Appendix \ref{app1}. The results are
presented in terms of Jacobi elliptic functions, rather than
theta-functions, as the Jacobi functions are more straightforward to handle
by means of computer-assisted algebra software. As an independent check we
verify that all the solutions given below satisfy Eqs. (\ref{cnlse}) by
direct differentiation with the software package Mathematica.

An important point is the choice of the theta function for $f$. The
functions $\vartheta _{1}(x)$ and $\vartheta _{2}(x)$ have real zeros,
therefore choosing them will give rise to singular solutions. The functions 
$\vartheta _{3}(x)$ and $\vartheta _{4}(x)$ are related by a phase shift of 
$\pi /2$. Previously known solutions \cite
{flo_trem_pla,kos_uzu_oc,flo_trem_oc} employ the choice $f=\vartheta
_{4}^{2}(x)$. In this work, we start with $f=\vartheta _{3}(x)\vartheta
_{4}(x)$, hence the results are different from the previously known ones.
Furthermore, the advantage of HM is that it can be readily generalized for
CNLSE with more than two components.

\begin{figure}[h]
\centerline{\scalebox{0.5}{\includegraphics{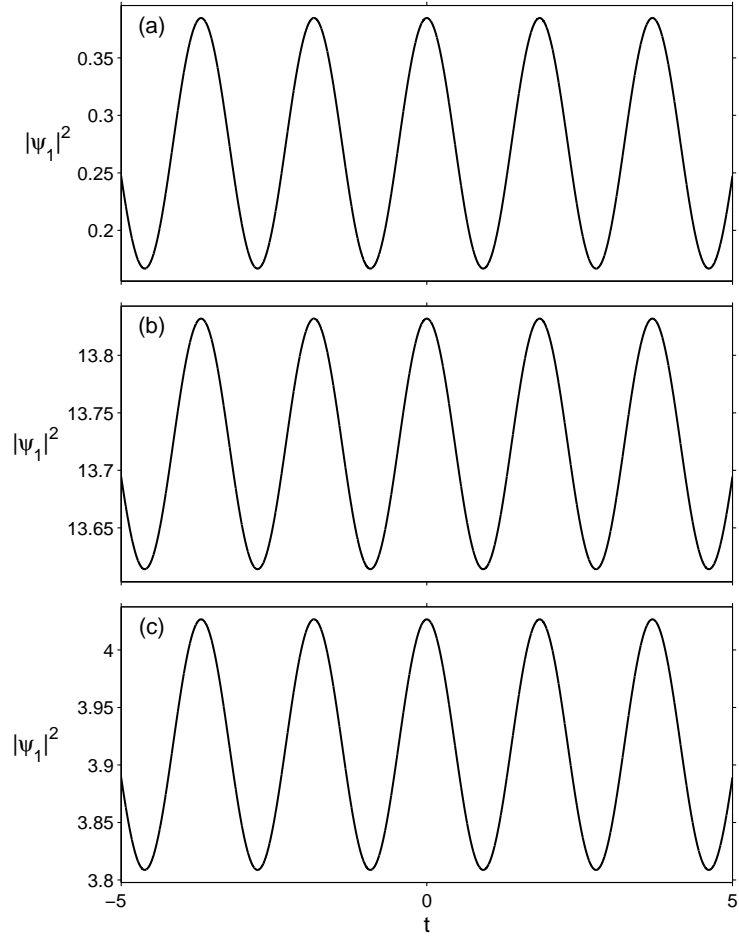}}}
\caption{ (a) The intensity $|\protect\psi_1|^2$ versus $t$ for the solution
(\ref{sol1}) with the amplitude relation (\ref{AB}) in the case $B_1 = 1$.
(b) The solution (\ref{sol2}) in the case $B_1 = 1$. (c) The solution (\ref
{sol3}) in the case $A_1 = A_2$. In all the three cases shown, and also
below in Fig. 2, $r = 1$ and $k =0.7$. Note the difference in the vertical
scale between the three solutions.}
\label{fig1}
\end{figure}

By means of this technique (see details in Appendix \ref{app1}), we have
obtained three new sets of solutions for the system (\ref{cnlsem}). The
first set has the form

\begin{eqnarray}
\psi _{1} &=&\frac{1}{2}\left\{ A_{1}\left[ \frac{(1-k^{2})^{1/4}}{{\rm {dn}}
(rt)}-\frac{{\rm {dn}}(rt)}{(1-k^{2})^{1/4}}\right] +\frac{ikB_{1}{\rm {sn}}
(rt){\rm cn}(rt)}{{\rm {dn}}(rt)}\right\} e^{-i\Omega z},  \nonumber \\
\psi _{2} &=&\frac{1}{2}\left\{ A_{1}\left[ \frac{(1-k^{2})^{1/4}}{{\rm {dn}}
(rt)}-\frac{{\rm {dn}}(rt)}{(1-k^{2})^{1/4}}\right] -\frac{ikB_{1}{\rm {sn}}
(rt){\rm cn}(rt)}{{\rm {dn}}(rt)}\right\} e^{-i\Omega z},  \label{sol1}
\end{eqnarray}

\noindent where $k$ is the modulus of the elliptic functions. The amplitude
parameters in these expressions are related by

\begin{equation}
B_{1}^{2}=\frac{k^{2}A_{1}^{2}}{\sqrt{1-k^{2}}}-\frac{8r^{2}k^{2}}{1+\sigma }
\,,  \label{AB}
\end{equation}
the frequency is

\begin{equation}
\Omega =r^{2}\sqrt{1-k^{2}}+\frac{3}{2}r^{2}(2-k^{2})-\frac{1}{4}
A_{1}^{2}(1+\sigma )\left[ \sqrt{1-k^{2}}+\frac{1}{\sqrt{1-k^{2}}}-2\right] ,
\label{Omega}
\end{equation}
and the modulus $k$ is related to the linear coupling constant $\kappa $:

\begin{equation}
\kappa =r^{2}\left( \sqrt{1-k^{2}}-1+\frac{1}{2}k^{2}\right) .
\label{kappa1}
\end{equation}
Thus, the solution (\ref{sol1}) contains two free parameters, as five
constants $A_{1},B_{1},k,r$, and $\Omega $ are subject to three constraints 
(\ref{AB}), (\ref{Omega}), and (\ref{kappa1}). Each of two other solutions
displayed below also contains five constants on which three conditions are
imposed. An example of the periodic solution (\ref{sol1}) is shown in Fig. 
\ref{fig1} (a).

The second solution is

\begin{eqnarray}
\psi _{1} &=&\frac{1}{2}\left\{ A_{1}\left[ \frac{(1-k^{2})^{1/4}}{{\rm {dn}}
(rt)}+\frac{{\rm {dn}}(rt)}{(1-k^{2})^{1/4}}\right] +\frac{ikB_{1}{\rm {sn}}
(rt){\rm cn}(rt)}{{\rm {dn}}(rt)}\right\} e^{-i\Omega z},  \nonumber \\
\psi _{2} &=&\frac{1}{2}\left\{ A_{1}\left[ \frac{(1-k^{2})^{1/4}}{{\rm {dn}}
(rt)}+\frac{{\rm {dn}}(rt)}{(1-k^{2})^{1/4}}\right] -\frac{ikB_{1}{\rm {sn}}
(rt){\rm cn}(rt)}{{\rm {dn}}(rt)}\right\} e^{-i\Omega z}.  \label{sol2}
\end{eqnarray}
The relation between the amplitude parameters in this solution is the same
as given by Eq. (\ref{AB}), but relations between other parameters are
different,

\[
\Omega =-r^{2}\sqrt{1-k^{2}}+\frac{3}{2}r^{2}(2-k^{2})-\frac{1}{4}
A_{1}^{2}(1+\sigma )\left[ \sqrt{1-k^{2}}+\frac{1}{\sqrt{1-k^{2}}}+2\right]
, 
\]

\begin{equation}
\kappa =-r^{2}\left( \sqrt{1-k^{2}}+1-\frac{1}{2}k^{2}\right) .
\label{kappa2}
\end{equation}
Figure \ref{fig1} (b) shows an example of this solution.

The third explicit solution is

\begin{eqnarray}
\psi _{1} &=&\frac{1}{2}\left\{ A_{1}\left[ \frac{(1-k^{2})^{1/4}}{{\rm {dn}}
(rt)}-\frac{{\rm {dn}}(rt)}{(1-k^{2})^{1/4}}\right] +iA_{2}\left[ \frac{
(1-k^{2})^{1/4}}{{\rm {dn}}(rt)}+\frac{{\rm {dn}}(rt)}{(1-k^{2})^{1/4}}
\right] \right\} e^{-i\Omega z},  \nonumber \\
\psi _{2} &=&\frac{1}{2}\left\{ A_{1}\left[ \frac{(1-k^{2})^{1/4}}{{\rm {dn}}
(rt)}-\frac{{\rm {dn}}(rt)}{(1-k^{2})^{1/4}}\right] -iA_{2}\left[ \frac{
(1-k^{2})^{1/4}}{{\rm {dn}}(rt)}+\frac{{\rm {dn}}(rt)}{(1-k^{2})^{1/4}}
\right] \right\} e^{-i\Omega z}.  \label{sol3}
\end{eqnarray}
This time, the amplitude parameters are related by the equation [cf. the
relation (\ref{AB}) for the two previous solutions]

\begin{equation}
A_{1}^{2}+A_{2}^{2}=\frac{8r^{2}\sqrt{1-k^{2}}}{1+\sigma }\,,  \label{AA}
\end{equation}
and other constants obey relations

\[
\Omega =-r^{2}(2-k^{2})-4r^{2}\sqrt{1-k^{2}}+(1+\sigma )A_{1}^{2}, 
\]
\begin{equation}
\kappa =2r^{2}\sqrt{1-k^{2}}.  \label{kappa3}
\end{equation}
Figure \ref{fig1} (c) shows an example of this periodic solution.

All the above solutions exist only in the presence of the linear coupling:
as it follows from Eqs. (\ref{kappa1}), (\ref{kappa2}), and (\ref{kappa3}),
no nontrivial solution can be found if $\kappa =0$. On the other hand, the
solutions exist at any value of $\sigma $, the Manakov's case, $\sigma =1$,
having nothing peculiar.

As it was mentioned above, the same HM-based procedure makes it possible to
reproduce all the other periodic solutions of the CNLSE system (\ref{cnlse})
which were reported earlier in Refs. \cite
{flo_trem_pla,kos_uzu_oc,flo_trem_oc}. Those solutions can be obtained on
the basis of different combinations of the theta functions, see details in
Appendix \ref{app2}.

\subsection{The normal-dispersion case}

Proceeding to Eqs. (\ref{cnlse}) with $\beta >0$, we adopt the normalization 
$\beta =1$, thus dealing with the equations

\begin{eqnarray}
i\psi _{1z}-\psi _{1tt}+\kappa \psi _{2}+\left( |\psi _{1}|^{2}+\sigma |\psi
_{2}|^{2}\right) \psi _{1} &=&0,  \nonumber \\
i\psi _{2z}-\psi _{2tt}+\kappa \psi _{1}+\left( |\psi _{2}|^{2}+\sigma |\psi
_{1}|^{2}\right) \psi _{2} &=&0.  \label{normal}
\end{eqnarray}

\noindent As well as in the case of the anomalous GVD considered above, we
can repeat the same steps, involving the transformation (\ref{ab}) and HM,
to generate periodic-wave solutions for the normal-GVD regime. We have thus
obtained two new sets of solutions, using different identities for
theta-functions.

The form of the first solution is the same as given by Eqs. (\ref{sol1}),
but the amplitude parameters are related by

\begin{eqnarray}  \label{AB1}
B_1^2 = \frac{k^2}{\sqrt{1-k^2}} \left(A_1^2 + \frac{8 r^2 \sqrt{1-k^2}} 
{1+\sigma}\right).
\end{eqnarray}

\noindent The relations (\ref{Omega}) and (\ref{kappa1}) for the solution 
(\ref{sol1}) are replaced by

\begin{eqnarray}
\Omega &=&-\frac{3}{2}r^{2}(2-k^{2})-r^{2}\sqrt{1-k^{2}}-\frac{1}{4}
(1+\sigma )A_{1}^{2}\left( \sqrt{1-k^{2}}+\frac{1}{\sqrt{1-k^{2}}}-2\right) ,
\label{nor-freq1} \\
\kappa &=&r^{2}\left( 1-\frac{1}{2}k^{2}-\sqrt{1-k^{2}}\right) .
\label{kappa_dark1}
\end{eqnarray}

\begin{figure}[h]
\centerline{\scalebox{0.5}{\includegraphics{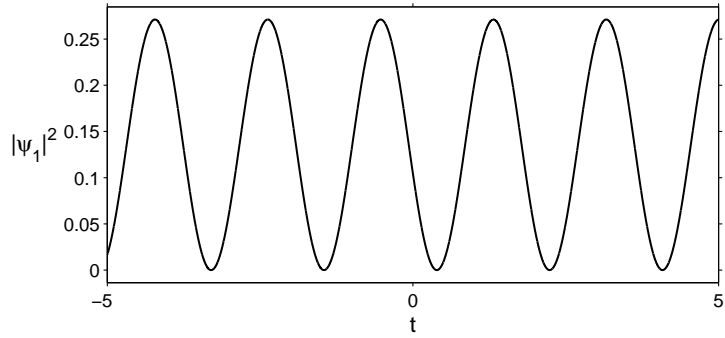}}}
\caption{ The intensity $|\protect\psi_1|^2$ versus $t$ for the solution 
(\ref{sol1}) of the dark-soliton type, subject to the amplitude relation (\ref
{AB}), in the case $B_1= 1$.}
\label{fig2}
\end{figure}

The form of the second solution is the same as given by Eq. (\ref{sol2}),
and the amplitude parameters are related by (\ref{AB1}). However, the other
parameters are related by

\begin{eqnarray}
\Omega &=&-\frac{3}{2}r^{2}(2-k^{2})+r^{2}\sqrt{1-k^{2}}-\frac{1}{4}
(1+\sigma )A_{1}^{2}\left( \sqrt{1-k^{2}}+\frac{1}{\sqrt{1-k^{2}}}+2\right) ,
\label{nor-freq2} \\
\kappa &=&r^{2}\left( \sqrt{1-k^{2}}+1-\frac{1}{2}k^{2}\right)
\label{kappa_dark2}
\end{eqnarray}
[cf. Eq. (\ref{kappa2})]. Figure \ref{fig2} shows an example of the latter
periodic solutions.

There is no counterpart for the third anomalous dispersion regime solution 
(\ref{sol3}) in the normal-GVD regime.

\section{The soliton limit}

A natural question is the soliton (long-wave) limit of the periodic
solutions derived above for both the anomalous- and normal-GVD regimes,
which corresponds to $k\rightarrow 1$. Note that many expressions in the
above solutions seem singular in this limit [for instance, Eq. (\ref{AA})].
However, accurate analysis shows that a nonsingular limit exists.

In the case of anomalous GVD, both solutions (\ref{sol1}) and (\ref{sol2})
give rise to a single long-wave limit,

\begin{eqnarray}
\psi _{1} &=&\frac{1}{2}\left[ \sqrt{\beta _{1}^{2}+\frac{8r^{2}}{1+\sigma }}
{\rm sech}\left( rt\right) +i\beta _{1}{\rm tanh}\left( rt\right) \right]
\exp \left( -i\Omega _{{\rm sol}}^{(1)}z\right) ,  \nonumber \\
\psi _{2} &=&\frac{1}{2}\left[ \sqrt{\beta _{1}^{2}+\frac{8r^{2}}{1+\sigma }}
{\rm sech}\left( rt\right) -i\beta _{1}{\rm tanh}\left( rt\right) \right]
\exp \left( -i\Omega _{{\rm sol}}^{(1)}z\right) ,  \label{soliton1}
\end{eqnarray}

\noindent where $\beta _{1}$ is an arbitrary real constant, and

\begin{equation}
\Omega _{{\rm sol}}^{(1)}=-\frac{1}{2}r^{2}-\frac{1}{4}(1+\sigma )\beta
_{1}^{2},\hspace{0.5in}\kappa =-\frac{1}{2}r^{2}.  \label{kappa}
\end{equation}

\noindent The solution (\ref{soliton1}) corresponds to a state in the form
of a bright soliton placed on top of a continuous-wave background, as it is
seen from the expression for the intensity:

\[
|\psi _{1}|^{2}=\left| \psi _{2}\right| ^{2}=\frac{1}{4}\left[ \beta
_{1}^{2}+\frac{8r^{2}}{1+\sigma }{\rm sech}^{2}\left( rt\right) \right] . 
\]
Solutions of this type are often called {\it anti-dark} solitons. \noindent

The long-wave limit of the solution (\ref{sol3}) is only possible, as it
follows from Eq. (\ref{kappa3}), if the linear coupling $\kappa $ vanishes
as $\sqrt{1-k^{2}}$ simultaneously with $\left( 1-k^{2}\right) $ [on the
contrary to the above soliton-limit solution, see Eq. (\ref{kappa})]. In the
corresponding limit form of the solution, both $\psi _{1}$ and $\psi _{2}$
are proportional to ${\rm sech}\left( rt\right) $, which implies an obvious
bright vector soliton.

In the case of normal GVD, the common long-wave limit of the solutions given
by Eqs. (\ref{sol1}), (\ref{AB1}), (\ref{nor-freq1}), (\ref{kappa_dark1})
and Eqs. (\ref{sol2}), (\ref{AB1}), (\ref{nor-freq2}), (\ref{kappa_dark2}) is

\begin{eqnarray*}
\psi _{1} &=&\frac{1}{2}\left[ -\beta _{2}{\rm sech}\left( rt\right) +i\sqrt{
\beta _{2}^{2}+\frac{8r^{2}}{1+\sigma }}{\rm tanh}\left( rt\right) \right]
\exp (-i\Omega _{{\rm sol}}^{(2)}z), \\
\psi _{2} &=&\frac{1}{2}\left[ -\beta _{2}{\rm sech}\left( rt\right) -i\sqrt{
\beta _{2}^{2}+\frac{8r^{2}}{1+\sigma }}{\rm tanh}\left( rt\right) \right]
\exp (-i\Omega _{{\rm sol}}^{(2)}z),
\end{eqnarray*}

\noindent where $\beta _{2}$ is an arbitrary real constant, and 
\[
\Omega _{{\rm sol}}^{(2)}=-\frac{3}{2}r^{2}-\frac{1}{4}(1+\sigma )\beta
_{2}^{2},\hspace{0.5in}\kappa =\frac{1}{2}r^{2}. 
\]

\noindent As expected, this is a dark soliton with the intensity distribution

\[
|\psi _{1}|^{2}=\frac{1}{4}\left[ \beta _{2}^{2}+\frac{8r^{2}}{1+\sigma}
\tanh ^{2}\left( rt\right) \right] . 
\]

\noindent In the general case ($\beta _{2}\neq 0$), this dark soliton is of
the {\it gray} type, i.e., with a nonzero minimum value of the intensity at
the central point, $t=0$.

\section{Conclusions}

The periodic solutions obtained in this work for the nonlinear fiber-optic
model with both anomalous and normal GVD can be helpful in predicting the
energy and width of ultra-short soliton trains using the bi-modal pulse
propagation in nonlinear optical fibers (the underlying CNLSE model is
definitely relevant for the pulse trains with the temporal width of the
pulses up to $\sim 1$ ps, provided that the loss is compensated by
appropriate gain). Using these analytical solutions, and knowing fiber
parameters, such as GVD, SPM and XPM coefficients, one can calculate the
exact shape, width and power of the periodic solution. These results may
also help to perform numerical simulations of the beatings of dual-frequency
signals.

The above consideration was not dealing with stability of the pulse trains.
In the case of the anomalous GVD, the trains may be unstable; however, in
many cases one may expect that their instability is quite weak (see, e.g.,
Ref. \cite{Arnold}), which makes it possible to consider practical
applications of these waves, especially in the case of limited propagation
distance, which will not give the instability enough room to develop. On the
other hand, the pulse train may be completely stable in the case of normal
GVD. Moreover, a stable wavetrain with the carrier frequency chosen in the
normal-dispersion spectral region can be used to stabilize a co-propagating
periodic wave launched in the anomalous-GVD region (on the other side of the
zero-dispersion point), provided that group velocities of both waves
coincide, as it was proposed in Ref. \cite{Ship}. In fact, it is a
challenging issue (which will be considered elsewhere) to find exact
solutions for periodic waves in the bimodal system, with the XPM coefficient 
$\sigma =2$ and without linear coupling ($\kappa =0$), in the case of
opposite signs of the dispersion in the two modes.

For completeness, we have also presented the long-wave limit of the newly
obtained periodic wave solutions, in both the anomalous- and normal-GVD
cases. In the former case, it turns out to be an anti-dark soliton, while in
the latter situation the long-wave limit amounts to a dark (generally, gray)
soliton.

To summarize, we have considered the CNLSE model (\ref{cnlse}), which
governs the propagation of electromagnetic waves in nonlinear bimodal
optical fibers. The model includes both XPM (nonlinear)\ and linear
couplings, the latter one accounting for a twist or elliptic deformation of
the fiber, in the cases of the linear and circular polarizations,
respectively. Using properties of elliptic theta functions and the Hirota
bilinear method, we have found three new types of periodic solutions in the
case of anomalous GVD, and two new types in the case of normal GVD. The new
solutions require the presence of the linear coupling. We have also shown
that all the previously reported periodic solutions of the CNLSE model can
also be obtained by means of the Hirota bilinear method. Thus, we conclude
that this method furnishes a powerful tool to derive different forms of
periodic-wave solutions in terms of elliptic functions. Basically the periodic solutions derived using the elliptic functions are in the form of raised sine or cosine functions, as seen from Figs. \ref{fig1} and \ref{fig2}. In most of the return-to-zero (RZ) coding the commonly used optical modulators utilize Mach-Zender interferometric technique. This technique yields raised cosine form of pulses for RZ bits coding. Hence we believe that the periodic solutions derived using Jacobi elliptic functions are of the same type as those generated by the Mach-Zender interferometer modulators. New analytical results obtained in this work may provide for important information in determining properties of pulse trains that are generated by dual-frequency signals in nonlinear optical fibers.

\begin{center}
{\bf ACKNOWLEDGEMENTS}
\end{center}

K.N. acknowledges support from the Research Grants Council (RGC) of the Hong
Kong Special Administrative Region, China (Project No. PolyU5132/99E). This
author is grateful to P.K.A. Wai for all kinds of help. K.W.C. acknowledges
a partial financial support from RGC contracts HKU 7066/00E and HKU
7006/02E. B.A.M. appreciates hospitality of the Optoelectronics Research
Centre at the Department of Electronics Engineering, the City University of
Hong Kong.

\newpage

\appendix

\section{The theta-functions identities}

\label{app1}

The theta functions $\theta _{n}(x)$, $n$ = 1,2,3,4, are defined by as
follows below, together with auxiliary parameters $q$ and $\tau $: 
\begin{eqnarray*}
\vartheta _{1}(x) &=&2\sum_{n=0}^{\infty }(-1)^{n}q^{(n+1/2)^{2}}\sin \left[
(2n+1)x\right]  \\
&\equiv &-\sum_{-\infty }^{\infty }\exp \left[ \pi i\tau \left( m+\frac{1}{2}
\right) ^{2}+2i\left( m+\frac{1}{2}\right) \left( x+\frac{\pi }{2}\right)
\right] ,
\end{eqnarray*}

\begin{eqnarray*}
\vartheta _{2}(x) &=&2\sum_{n=0}^{\infty }q^{(n+1/2)^{2}}\cos \left[
(2n+1)x\right]  \\
&\equiv &\sum_{-\infty }^{\infty }\exp \left[ \pi i\tau \left( m+\frac{1}{2}
\right) ^{2}+2i\left( m+\frac{1}{2}\right) x\right] ,
\end{eqnarray*}

\begin{eqnarray*}
\vartheta _{3}(x) &=&1+2\sum_{n=1}^{\infty }q^{n^{2}}\cos \left( 2nx\right) 
\\
&=&\sum_{-\infty }^{\infty }\exp (\pi i\tau m^{2}+2imx),
\end{eqnarray*}

\begin{eqnarray*}
\vartheta _{4}(x) &=&1+2\sum_{n=1}^{\infty }(-1)^{n}q^{n^{2}}\cos \left(
2nx\right)  \\
&=&\sum_{-\infty }^{\infty }\exp \left[ \pi i\tau m^{2}+2im\left( x+\frac{
\pi }{2}\right) \right] ,
\end{eqnarray*}
\begin{equation}
q=\exp (\pi i\tau )\,.  \label{q}
\end{equation}
Note that, in the present case, $\tau $ is imaginary, hence the constant $q$
defined in Eq. (\ref{q}) belongs to the interval $0<q<1$. It is related to
the modulus $k$ of the corresponding Jacobi functions as $q=\exp \left[ -\pi
K\left( \sqrt{1-k^{2}}\right) /K(k)\right] $, where $K$ is the complete
elliptic integral of the first kind.  Zeros of the functions $\theta _{1}$, 
$\theta _{2}$, $\theta _{3}$, and $\theta _{4}$ are at the points $x=M\pi
+N\pi \tau $, $(M+1/2)\pi +N\pi \tau $, $(M+1/2)\pi +(N+1/2)\pi \tau $, and 
$M\pi +(N+1/2)\pi \tau $, respectively, where $M$ and $N$ are integers.

There are many identities involving products of theta functions which are
relevant to the Hirota bilinear method. An example is

\begin{equation}
\vartheta _{4}(x+y)\vartheta _{4}(x-y)\vartheta _{2}^{2}(0)=\vartheta
_{4}^{2}(x)\vartheta _{2}^{2}(y)+\vartheta _{3}^{2}(x)\vartheta _{1}^{2}(y).
\label{y}
\end{equation}

\noindent A proof can be found in Ref. \cite{kwc2}. On differentiating Eq. 
(\ref{y}) with respect to $y$ twice and setting $y=0$, one obtains

\[
D_{x}^{2}\vartheta _{4}(x)\cdot \vartheta _{4}(x)=2\vartheta
_{3}^{2}(0)\vartheta _{4}^{2}(0)\vartheta _{3}^{2}(x)+\frac{2\vartheta
_{2}^{\prime \prime }(0)\vartheta _{4}^{2}(x)}{\vartheta _{2}(0)},
\]
where the Hirota derivative [see Eq. (\ref{hir-der})] is

\begin{eqnarray}
D_x^2 g \cdot f = f g_{xx} - 2 f_x g_x + g f_{xx}.
\end{eqnarray}
From these results, one can construct equations of increasing complexity by
using Hirota identities like

\[
D_{x}^{2}ab\cdot cd=bd(D_{x}^{2}a\cdot c)+ac(D_{x}^{2}b\cdot
d)+2(D_{x}a\cdot c)(D_{x}b\cdot d).
\]
The novel choice in the present paper is to set $f=\vartheta
_{3}(x)\vartheta _{4}(x)$ as opposed to the selection of $f=\vartheta
_{4}^{2}(x)$ used implicitly in earlier works (see Appendix \ref{app2}). The
relevant identity is then

\begin{eqnarray*}
D_{x}^{2}f\cdot f &=&D_{x}^{2}\vartheta _{3}(x)\vartheta _{4}(x)\cdot
\vartheta _{3}(x)\vartheta _{4}(x) \\
&=&2\vartheta _{3}^{2}(0)\vartheta _{4}^{2}(0)\left[ \vartheta
_{3}^{4}(x)+\vartheta _{4}^{4}(x)\right] +\frac{2\vartheta _{2}^{\prime
\prime }(0)\vartheta _{3}^{2}(x)\vartheta _{4}^{2}(x)}{\vartheta _{2}(0)}.
\end{eqnarray*}
The trick is then to select the amplitude parameters such that only the term 
$\vartheta _{3}^{2}(x)\vartheta _{4}^{2}(x)$ remains in the square bracket
of Eq. (\ref{trans3}). The remaining terms of the trilinear equations are
then satisfied by adjusting the frequency parameter $\Omega $.

%\newpage

\section{Comparison with previously known results}

\label{app2}

Previously known periodic solutions of the CNLSE model (\ref{cnlse}) can be
recovered by the present approach via proper choice of theta functions in
Eqs. (\ref{trans3}). For instance, selecting

\begin{eqnarray*}
f &=&\theta _{4}^{2}(\alpha t), \\
g &=&A_{0}\theta _{1}(\alpha t)\theta _{2}(\alpha t),\hspace{0.25in}
A_{0}^{2}=\frac{24\alpha ^{2}\vartheta _{2}^{4}(0)\vartheta _{4}^{2}(0)}
{(1+\sigma )\vartheta _{3}^{2}(0)}, \\
G &=&B_{0}\theta _{2}(\alpha t)\theta _{3}(\alpha t),\hspace{0.25in}
B_{0}^{2}=\frac{24\alpha ^{2}\vartheta _{2}^{2}(0)\vartheta _{4}^{4}(0)}
{(1+\sigma )\vartheta _{3}^{2}(0)},
\end{eqnarray*}
reproduces a solution reported in Ref. \cite{kos_uzu_oc}, which takes the
following form in the present notation:

\begin{eqnarray}
a &=&2\sqrt{\frac{6}{1+\sigma }}rk^{2}\,{\rm {sn}}(rt)\,{\rm cn}(rt)\exp
(-i\Omega z),  \label{eq:app_b1} \\
b &=&2\sqrt{\frac{6}{1+\sigma }}rk\,{\rm {cn}}(rt)\,{\rm dn}(rt)\exp
(-i\Omega z),  \label{eq:app_b2}
\end{eqnarray}

\[
\Omega =\frac{5}{2}r^{2}(1-2k^{2}),\,\,\kappa =\frac{3}{2}r^{2}.
\]
$\alpha$ is the wavenumber in the theta function representation. 
Since arguments of theta and elliptic functions differ by a scalar
(Eq. (B3) below), the wavenumbers in the two formulations
are related by $r = \alpha \theta_3^2(0)$. The Jacobi elliptic functions appearing in the expressions (\ref{eq:app_b1})
and (\ref{eq:app_b2}) are related to the theta functions as follows:

\begin{eqnarray}
{\rm {sn}}(rt) &=&\frac{\vartheta _{3}(0)\vartheta _{1}(x)}{\vartheta
_{2}(0)\vartheta _{4}(x)},\hspace{0.5in}{\rm {cn}}(rt)=\frac{\vartheta
_{4}(0)\vartheta _{2}(x)}{\vartheta _{2}(0)\vartheta _{4}(x)}, \nonumber \\
{\rm {dn}}(rt) &=&\frac{\vartheta _{4}(0)\vartheta _{3}(x)}{\vartheta
_{3}(0)\vartheta _{4}(x)},\hspace{0.5in}rt\equiv \frac{x}{\vartheta
_{3}^{2}(0)}.
\end{eqnarray}
Note a difference in the scale factors in the arguments of the two types of
elliptic functions.


\begin{references}
\bibitem{yamamoto00}  T.\ Yamamoto, E.\ Yoshida, K.\ R.\ Tamura, K.\
Yonenaga and N.\ Nakazawa, IEEE\ Photon.\ Technol.\ Lett.\thinspace\ 
{\bf 12}, 353 (2000).

\bibitem{hasegawa84}  A.\ Hasegawa, Opt.\ Lett.\thinspace\ {\bf 9}, 288
(1984).

\bibitem{tai86}  K.\ Tai, A.\ Tomita, J.\ L.\ Jewell and A.\ Hasegawa, Appl.
Phys. Lett.\thinspace\ {\bf 49}, 236 (1986).

\bibitem{mamyshev91}  P.\ V.\ Mamyshev, S.\ V.\ Chernikov and E.\ M.\
Dianov, IEEE J. Quantum Electron.\thinspace\ {\bf 27}, 2347 (1991).

\bibitem{chernikov93}  S.\ V.\ Chernikov, D.\ J.\ Richardson, R.\ I.\
Laming, E.\ M.\ Dianov and D.\ N.\ Payne, Appl. Phys. Lett.\thinspace\ {\bf
63}, 293 (1993).

\bibitem{chernikov93_2}  S.\ V.\ Chernikov and J.\ R.\ Taylor, Electron.
Lett.\thinspace\ {\bf 29}, 658 (1993).

\bibitem{chernikov93_3}  S.\ V.\ Chernikov, J.\ R.\ Taylor and R.\ Kashyap,
Electron. Lett.\thinspace\ {\bf 29}, 1788 (1993).

\bibitem{chernikov94}  S.\ V.\ Chernikov, J.\ R.\ Taylor and R.\ Kashyap,
Electron. Lett.\thinspace\ {\bf 30}, 433 (1994).

\bibitem{chernikov94_2}  S.\ V.\ Chernikov, J.\ R.\ Taylor and R.\ Kashyap,
Opt. Lett.\thinspace\ {\bf 19}, 539 (1994).

\bibitem{chernikov96}  S.\ V.\ Chernikov, R.\ Kashyap, M.\ J.\ Guy, D.\ G.\
Moodie and J.\ R.\ Taylor, Phil. Trans. R. Soc. Lond. A\thinspace\ 
{\bf 354}, 719 (1996).

\bibitem{swanson95}  E.\ A.\ Swanson and S.\ R.\ Chinn, IEEE Photon.
Technol. Lett.\thinspace\ {\bf 7}, 114 (1995).

\bibitem{trillo94}  S.\ Trillo, S.\ Wabnitz and T.\ A.\ B.\ Kennedy, Phys.\
Rev.\ A\thinspace\ {\bf 50}, 1732 (1994).

\bibitem{pitois02}  S.\ Pitois, J.\ Fatome and G.\ Millot, Opt.\
Lett.\thinspace\ {\bf 27}, 1729 (2002).

\bibitem{HAS-KOD}  A.\ Hasegawa and Y.\ Kodama, {\em Solitons in Optical
Communication} (Oxford University Press, New York, 1995).

\bibitem{GPA}  G.\ P. Agrawal, {\em Nonlinear Fiber Optics} (Academic Press,
San Diego, 1989).

\bibitem{MAN}  S.\ V.\ Manakov, Sov.\ Phys.\ JETP\ {\bf 38}, 248 (1974).

\bibitem{Hiet}  J. Hietarinta, Phys. Lett. A 96, 273 (1983).

\bibitem{men}  C.\ R.\ Menyuk, IEEE\ J.\ Quant.\ Electron.\ {\bf 25}, 2674
(1989).

\bibitem{pka}  C.\ J.\ Chen, P.\ K.\ A.\ Wai and C.\ R.\ Menyuk, Opt.\ Lett. 
{\bf 15}, 479 (1990).

\bibitem{hir}  R.\ Hirota, J.\ Math.\ Phys.\thinspace\ {\bf 14}, 805 (1973).

\bibitem{flo_trem_pla}  M.\ Florja\'{n}czyk and R.\ Tremblay, Phys.\ Lett.\
A\thinspace\ {\bf 141}, 34 (1989).

\bibitem{kos_uzu_oc}  N.\ A. Kostov and I.\ M.\ Uzunov, Opt.\
Commun.\thinspace\ {\bf 89}, 389 (1992).

\bibitem{flo_trem_oc}  M.\ Florja\'{n}czyk and R.\ Tremblay, Opt.\
Commun.\thinspace\ {\bf 109}, 405 (1994).

\bibitem{kwc1}  K.\ W.\ Chow, J.\ Phys.\ Soc.\ Japan\thinspace\ {\bf 69},
1313 (2000).

\bibitem{kwc2}  K.\ W.\ Chow, Phys.\ Lett.\ A\thinspace\ {\bf 179}, 299
(1993).

\bibitem{Arnold}  J.M. Arnold, A.D. Boardman, H.M. Mehta, and R.C.J. Putman,
Opt. Commun. {\bf 122}, 48 (1995).

\bibitem{Ship}  A. Shipulin, G. Onishchukov, and B.A. Malomed, J. Opt. Soc.
Am. B {\bf 14}, 3393 (1997).
\end{references}
\end{document}